\documentclass[12pt,preprint]{aastex}

\usepackage{titletoc}
\usepackage{graphicx, amsfonts, amsmath, amssymb, amscd,  theorem, exscale,
epic, eepic, epsfig}

\usepackage{hyperref} %

\usepackage{makeidx}
\makeindex

\newcounter{Figure}
\theoremheaderfont{\scshape}

\theorembodyfont{\upshape}

\theoremheaderfont{\scshape\small} \theorembodyfont{\scshape\small}

\newcommand{\slap}{\mbox{$ \triangle \mkern -13mu / \ $}}
\newcommand{\nlap}{\mbox{$ \nabla \mkern -13mu / \ $}}
\renewcommand{\dlap}{\mbox{$ div \mkern -13mu / \ $}}

\newcommand{\be}{\begin{equation}}
\newcommand{\ee}{\end{equation}}
\newcommand{\bea}{\begin{eqnarray}}
\newcommand{\eea}{\end{eqnarray}}
\newcommand{\beas}{\begin{eqnarray*}}
\newcommand{\eeas}{\end{eqnarray*}}

\renewcommand{\sun}{\mbox{$ \circ \mkern -7mu   \cdot \  $}}

\shorttitle{The Electromagnetic Christodoulou Effect}
\begin{document}
\vspace{5pt}
\begin{center}
{\large \bf The Electromagnetic Christodoulou Memory Effect and its Application to Neutron Star Binary Mergers } 
\end{center}
\vspace{5pt}
\begin{center}
\footnote{ 
Lydia Bieri, University of Michigan, Department of Mathematics, Ann Arbor MI.
lbieri@umich.edu \\ 
PoNing Chen, Harvard University, Department of Mathematics, Cambridge MA.
pchen@math.harvard.edu \\ 
Shing-Tung Yau, Harvard University, Department of Mathematics, Cambridge MA.
yau@math.harvard.edu}
%
{\large \bf Lydia Bieri, PoNing Chen, Shing-Tung Yau } 
\end{center}
%
%
%
\noindent

\begin{abstract}
Gravitational waves are predicted by the general theory of relativity. 
It has been shown that gravitational waves have a
nonlinear memory, displacing test masses permanently. 
This is called the Christodoulou memory. 
We proved that the electromagnetic field contributes at highest
order to the nonlinear memory effect of gravitational waves, enlarging the permanent displacement of test masses.  
In experiments like LISA or LIGO which measure distances of test masses, the Christodoulou memory will manifest itself as a permanent displacement of these objects. 
It has been suggested to detect the Christodoulou memory effect using radio telescopes 
investigating small changes in pulsar's pulse arrival times. 
The latter experiments are based on present-day technology and measure changes in frequency. 
In the present paper, we study the electromagnetic Christodoulou memory effect and
compute it for binary neutron star mergers. These are typical sources of
gravitational radiation. 
During these processes, not only mass and momenta are radiated away in form of 
gravitational waves, but also very strong magnetic fields are produced and
radiated away. Moreover, a large portion of the energy is carried away by neutrinos. 
We give constraints on the conditions, where the energy transported by electromagnetic radiation is of similar or slightly higher order than the energy radiated in gravitational waves or in form of neutrinos. 
We find that for coalescing neutron stars, large magnetic fields magnify the Christodoulou memory as long as 
the gaseous environment is sufficiently rarefied. 
Thus the observed effect on test masses of a laser interferometer gravitational wave
detector will be enlarged by the contribution 
of the electromagnetic field. Therefore, the present results are important for the
planned experiments. 
Looking at the null asymptotics of spacetimes, which are solutions of the
Einstein-Maxwell equations, we derive the electromagnetic
Christodoulou memory effect. 
We obtain an exact solution of the full nonlinear problem, no approximations were used. 
Moreover, our results allow to answer astrophysical questions, as 
the knowledge about the amount of energy radiated away in a neutron star binary
merger enables us to gain information about the source of the gravitational waves. 
\end{abstract}
\section{Introduction} \label{intro}
The main goal of this paper is to discuss the electromagnetic Christodoulou memory
effect of gravitational waves and to compute 
this effect for typical sources. 
\cite{chrmemory} showed that gravitational waves have a
nonlinear memory, displacing test masses permanently. 
\cite{lpst1} proved 
that for spacetimes solving the Einstein-Maxwell (EM) equations, the
electromagnetic field 
contributes at highest order \footnote{Here, the word `order' refers to decay behavior of the exact solution, not to any approximations. That is, `higher order' means `less decay'. For details, see \cite{lpst1}. \cite{lydia2}, \cite{zip2}.}
 to the nonlinear memory effect of gravitational waves. 
In the present paper, we calculate it for neutron star binary mergers. 
We find
that for binary neutron star (BNS) mergers which occur in regions where the gaseous environment is sufficiently rarefied, 
very strong magnetic fields enlarge the Christodoulou memory effect considerably. 
So far fields which are strong enough to have a detectable effect have only been found to be generated during
coalescence of neutron star binaries or in magnetars. 

The electromagnetic Christodoulou memory cannot be considered as a `purely electromagnetic effect' nor as some `purely electromagnetic effect in neutron star binaries'. But it is an effect emerging from the Einstein-Maxwell equations (\ref{EM}) and as such it is indeed a gravitational-electromagnetic effect. 
Through the Einstein-Maxwell equations, the electromagnetic field enters the system of partial differential equations, and is therefore intrinsically connected with what we are used to call `gravitational part'. 
We show that to the nonlinear memory effect of gravitational waves found by \cite{chrmemory} there is a contribution from the electromagnetic field. 
We call this the electromagnetic Christodoulou effect (\cite{lpst1}). 

Our results emerge from our studies of  the Einstein-Maxwell equations (\ref{EM}), which describe the general relativistic theory of gravitation and electromagnetism. 
We show that the electromagnetic field on the right hand side of Einstein-Maxwell equations (\ref{EM}) generates curvature of the spacetime contributing to the Christodoulou effect. Our new result could be observed in gravitational wave experiments explained in section \ref{gravitationalwaveexp} with non-charged test masses. 

At this point, we would like to mention that a `memory of gravitational waves' had been known in a `linear' theory for a long time, but it was thought to be 
`negligible'. This is for instance explained in \cite{chrmemory}, and we do not want to go into these details here. Then \cite{chrmemory} showed that the 
`memory of gravitational waves' is really a nonlinear effect and larger than expected. Whereas his paper treats the Einstein vacuum equations, he mentions on p. 1488 how matter (i.e. electromagnetic or neutrino) radiation could contribute to this effect. It remained open, what the structures of these terms are, in particular what the asymptotics look like. In order to find out if a contribution to the nonlinear memory exists and how large it is, one has to solve the 
full nonlinear problem. In case of electromagnetic fields, we did that \cite{lpst1}. We shall come back to this point in detail  
at the end of section \ref{gravitationalwaveexp}. 
Moreover, we would like to point out that \cite{re1} in an interesting paper looked at situations with neutrino radiation for weak gravitational fields and derived a formula for the 
linearized problem. 
This is a very nice step towards the right direction. 
However, the linearized theory is unphysical and even contradicts Newtonian physics as we are going to explain in the following paragraphs. 
Linearized theory might give some hint that there might be a contribution to the memory effect from neutrino radiation, but it is unable to compute it. 
General relativity (GR) being an extension of Newtonian theory, the latter has to emerge from GR theory as a limit. 
Therefore what needs to be done in order to decide 
whether neutrino radiation contributes to the nonlinear memory, is to compute the full nonlinear picture. Only this will reveal the physics. 

The electromagnetic Christodoulou memory effect of gravitational waves is a nonlinear effect, and as such it is a manifestation of one of the very deep properties of the Einstein-Maxwell equations. This point requires further explanation: The Einstein equations where the right hand side is zero are highly nonlinear (in (\ref{EM}), set in the first equation the right hand side equal to zero),  whereas the Maxwell equations are linear. However, when we plug in on the right hand side of the Einstein equations the stress-energy tensor of an electromagnetic field and complement the Einstein system to the full Einstein-Maxwell equations (\ref{EM}) (EM), this system of partial differential equations is highly nonlinear. 
Here, we are facing a fundamental and very deep property of the EM equations: Even though electromagnetic theory by itself is a linear theory, the coupled Einstein-Maxwell system is nonlinear. We cannot `split' the two theories anymore. When we work with electromagnetic fields in general relativity, we have to consider the full nonlinear theory of the Einstein-Maxwell equations. 

On the one hand, the EM equations exhibit linear effects, which one hopes to observe. The instantaneous displacements of test masses correspond to such linear effects. On the other hand, the EM equations feature nonlinear effects. The electromagnetic Christodoulou memory effect is such a nonlinear effect. 

Let us explain the latter point in more details. 
Our result is exact and stays valid for large data. In particular, in our paper \cite{lpst1} we obtained an exact solution for the nonlinear memory effect with contribution from the electromagnetic field, using geometric-analytic methods. We did not use any approximation method. Our result being exact, it is valid not only for small but also for large data, which describe 
typical sources for gravitational waves with memory. 
Thus, our exact formula is different from anything one might obtain using approximation methods. If one linearizes the Einstein-Maxwell system around Minkowski spacetime and uses the method of approximations to compute a memory effect from the stress-energy tensor (what some people call a `linear' effect, and what would correspond to a limit in weak gravity) one gets a formula which looks like our formula, but which is in fact different. No matter to what order one computes the terms in the infinite series, one will not get an exact result.
Using the approximation method, one could take the linear part or higher orders in the expansion into account or even compute a result to all orders, but still one has to make sure that the infinite series converges. That's where the constraints come into play and the results based on approximation methods only allow you to have small deviations from flat space. 
Meaning that approximation methods do not capture the physics, whereas our rigorous result gives the full physical picture. 
\cite{dcbs} showed that in general the corresponding series does not converge even for small energy-momentum tensor. 
More precisely, they first point out that in the Einstein equations one cannot prescribe the energy-momentum tensor as a given field 
because of the divergence property. However, if the harmonic condition is imposed one obtains the well-known reduced Einstein equations 
$R_{\mu \nu}^h = \epsilon( T_{\mu \nu} - \frac{1}{2} T g_{\mu \nu} )$. We inserted on the right hand side a small positive $\epsilon$. In the general case, where $\epsilon = 1$, this equation has solutions for arbitrary prescribed right hand side. But these solutions only solve the original Einstein field equations if the constraint equations are fulfilled and if the divergenc-free property holds. However, this last property cannot be realized  with respect to the metric $g_{\mu \nu}$. Instead in approximations the flat Minkowski metric is used with respect to which the divergence of the energy-momentum tensor vanishes. \cite{dcbs} show that for each $\epsilon>0$ there is a $n(\epsilon)$ 
such that, as the order of approximation increases to $n(\epsilon)$ the error decreases, achieving 
a minimum at order $n(\epsilon)$. However if the order of approximation is increased further beyond 
$n(\epsilon)$, the error again increases. This $n(\epsilon)$ tends to infinity as $\epsilon$ tends to zero. 
Thus for any given $\epsilon>0$ there is no convergence, in fact the sequence of approximants diverges 
as the order tends to infinity. On the other hand, for any fixed order of approximation the error 
tends to zero as $\epsilon$ tends to zero. 
It then trivially follows that this method completely fails for large data. 

The method we used in \cite{lpst1} is the same as \cite{chrmemory} introduced in his pioneering paper. He did not use any approximations, but geometric-analytic methods from which he obtained an exact formula. 

Moreover, there is the following problem with the method of approximations: 
Let us consider the Einstein equations with arbitrary energy-momentum tensor. For instance in the setting of a perfect fluid, we insert pressure, density and four-velocity of the fluid on the right hand side, in the electromagnetic case the electromagnetic field enters.  
Then linearize the equations around Minkowski space. This implies the following: The divergence of the energy-momentum tensor with respect to the flat metric is zero. This means that the influence of gravitation of motion of matter is ignored. In case of a perfect fluid, one obtains relativistic Euler equations without gravitation. One has motion without gravitation. In fact, this does not contain the Newtonian theory at all. Now, what does this mean, if the Newtonian theory is not present as a limit?  

In contrast to the problem of approximations just explained, our method takes care of the full nonlinear problem and we obtain a result which is no approximation but exact and which is true in all situations including large data. That is where all the approximation methods fail. 

We apply our new result on the electromagnetic Christodoulou memory to neutron star binary mergers. Our new effect is likely to be seen in gravitational wave experiments with neutron star coalescence as sources. 

Applying our new and exact formula to astrophysical data in the present paper, we give rough estimates. It will be a lot of interesting work for the future, to investigate all the different astrophysical scenarios in details and to compute the corresponding precise pictures. 
We are making available here our new method and exact formula to other scientists, especially to astrophysicists, and invite them to combine their knowledge with our new formula and to investigate the many scenarios in a precise way. 

Typical sources for gravitational waves are binary neutron star mergers and binary
black hole mergers. During such processes mass and momenta
are radiated away. Large magnetic fields are produced and radiated when the objects are neutron stars. The radiation travels at the speed of light. Geometrically this corresponds to the radiation 
moving along null hypersurfaces of corresponding spacetimes. 
Therefore it is important to
investigate these spacetimes.  
As the sources are very far away, we can think of us as doing the experiment at null
infinity, that is at the limit $t \to \infty$ of these null hypersurfaces. 
Thus it is crucial to understand the geometry of spacetimes
especially at null infinity, that is when $t \to \infty$ along null
hypersurfaces.  

The observation of gravitational waves not only will yield another proof of Einstein's theory of general relativity, but it will also 
open a new window for astronomical observations. Combined with other astrophysical observations, gravitational wave data will enhance understanding of energetic phenomena such as gamma ray bursts. Allowing to `see' deeper into formerly impenetrable regions, it will lay open these parts of the universe to us. Here, the Christodoulou memory effect provides a unique signature for detection as well as important information about the different forms of energy radiated in the gravitational wave source. 
Detailed knowledge about the amount of energy as well as the fractions of energy  
radiated through gravitational waves, electromagnetic waves and neutrinos in BNS mergers will shed light on the physics of the sources. 

It has been suggested to detect the Christodoulou memory effect using radio telescopes 
pointing at pulsars which serve as precise clocks. Their rapid rotation being extremely stable, 
one wants to analyze the pulse arrival times from a set of pulsars. 
Gravitational waves traveling by are seen as fluctuations in spacetime, 
they 
modulate the Earth-pulsar distance, which changes the observed time of arrivals of pulses. 
Pulsar timing arrays are most sensitive to supermassive black hole binary mergers.  
These experiments are based on present-day technology. For further details and references, we refer to: (\cite{fakir1}), (\cite{fakir2}), (\cite{fakir3}), (\cite{fakir4}), (\cite{favata9}), (\cite{favata}), 
(\cite{pshirkov1}), (\cite{seto1}), (\cite{vanhaasteren1}). 
Further, the Einstein Telescope will be sensitive to many more scenarios including binary neutron star mergers. 
These experiments together with LISA and Earth-based experiments like LIGO etc. are also expected to become important in astrophysical research. 
It is crucial to understand when electromagnetic energy contributes to observed signals and to the memory effect, and how large this contribution is. 

The results of \cite{lpst1} and the present paper answer the fundamental question about the role of magnetic fields played in the Christodoulou memory and in BNS mergers. We also investigate BNS coalescence that occur in media like gas rich or gas poor galaxies and in the intergalactic medium. 
We analyze signals from different regions and give constraints on plasma frequency, thus electron density of surrounding material. 

A major goal of general relativity (GR) and astrophysics is to precisely describe
and finally observe gravitational radiation, one of the 
predictions of GR. We know from the work \cite{chrmemory} that also
these waves radiate. 
In a laser interferometer gravitational wave detector, this will show in a
permanent displacement 
of test masses after a wave train has passed. 
The latter is known as the 
Christodoulou nonlinear memory effect. 
See also (\cite{payne}), (\cite{thorne}), (\cite{blanchetdamour}), (\cite{wiswil1}).  
\cite{chrmemory} showed how to measure the nonlinear memory effect 
as a permanent displacement of test
masses in such detectors. He derived
a precise formula for the permanent displacement in the Einstein vacuum (EV) case. 
$\Sigma^+ - \Sigma^-$ governs this permanent displacement. 
$\Sigma$ denotes the asymptotic shear of outgoing null hypersurfaces $C_u$
that are level sets of a foliation of the spacetime by an 
optical function $u$. $\Sigma^+$ and $\Sigma^-$ are the
limits of $\Sigma$ as 
$u$ tends to $+ \infty$ respectively $- \infty$. 

In \cite{lpst1}, we study the permanent displacement formula
for uncharged test particles of the same gravitational-wave experiment
for EM equations. 
We derive $\Sigma^+ - \Sigma^-$ in the
EM case finding that the electromagnetic field
changes the leading order term.  \footnote{Here, the word `order' refers to decay behavior of the exact solution, not to any approximations. That is, `higher order' means `less decay'. For details, see \cite{lpst1}. \cite{lydia2}, \cite{zip2}.} 
We show 
that the electromagnetic field does not enter the leading
order term of the Jacobi equation. As a result, to leading order, it does not change
the instantaneous displacement of test particles. 

This article is organized as follows. Section 2 explores neutron stars with magnetic fields and their importance for the Christodoulou memory. 
In section 3 we compute the electromagnetic Christodoulou memory effect. Section 4 shows how these results relate to gravitational wave experiments. Finally, in section 5 we investigate binary neutron star mergers.

\section{Neutron stars and magnetic fields} \label{ns}

As the most important sources of gravitational waves - besides binary neutron star mergers - figure binary black hole coalescence,  mergers of a black hole and a neutron star, supernova collapse and short gamma ray bursts (GRB). Among the current models for (short) GRB, the most promising from our point of view (large electromagnetic fields) are the ones powered by neutron star mergers or magnetars. See the recent overview of short GRB (\cite{berger}). 
In what follows, we are going to discuss sources of gravitational waves with large electromagnetic fields in view of the Christodoulou memory. For our purposes, BNS mergers will provide the best data. 

BNS mergers are frequent events, during which the stars radiate away energy. They are losing a certain amount of their original mass by the emission of gravitational waves, neutrinos and electromagnetic waves. One would expect to find a memory effect for all of these three forms of radiation. Whereas the pioneering paper of \cite{chrmemory} treated the purely gravitational memory, we are considering the electromagnetic case, for the neutrinos the question remains still open. 

The following questions have to be answered for the electromagnetic Christodoulou memory from BNS mergers: 
Under which physical conditions is the fraction of the energy carried away by electromagnetic waves big enough to considerably enlarge the purely gravitational memory? 
Where do we find the BNS that fulfill point one? 
What are the ratios of energy radiated away in a merger of such BNS? 

Here, we are interested to give some rough estimates in view of the application of our result to BNS mergers. We leave the details to the specialists who do simulations of such mergers. In the present paper, we investigate the situations, where the electromagnetic field plays an important role, give constraints on energies radiated away, and conclude with cases, where the electromagnetic Christodoulou memory effect can be observed in experiments. 

In a merger of two neutron stars into a black hole, the frequency of the electromagnetic wave which would carry away the magnetic energy, should be of about the order of the inverse of the light travel time for a neutron star radius: Assuming a neutron star radius of $10 km$, this yields 
$\frac{300'000 km/sec}{  10km}$, that is $30 kHz$. Scaling with the orbital frequency before merger gives $4 kHz$. 
This is of the same order as the plasma frequency of $60 rad/s$ respectively $\approx 9 kHz$ for interstellar medium, implying that  in this case most of the energy of the electromagnetic wave would be dispersed or absorbed in the surrounding medium. 

The plasma frequency $\omega_p$ of a medium with free electron density $n$ is given by:
\be \label{plasmafreq1}
\omega_p^2=\frac{4\pi n e^2}{m}
\ee
where $m$ and $e$ is the electron mass and charge respectively. 
For interstellar regions (spiral arms), the electron density is about $ 1 cm^{-3}$. It may be $0.1 cm^{-3}$ in between. 

On the other hand, the amount of energy carried by high frequency electromagnetic radiation
in the coalescence of BNS  is not expected to
exceed $10^{51}$ ergs. This is the maximum energy output that occurs in
short gamma-ray bursts (GRBs). 

During a BNS merger, tidal gravitational forces induce enormous heat production. This heat will largely be radiated in 
neutrinos before a black hole forms. According to \cite{dessart} energies radiated by neutrinos reach 
$~ 10^{53} ergs$.  The following papers treated neutrino radiation in similar situations: 
 (\cite{ruffert}), (\cite{ruffert2}), (\cite{rossw2}), (\cite{rossw3}), (\cite{seti}), (\cite{seti2}). 
The immediate question arises: Will the energy loss by neutrinos dominate the energy loss due to electromagnetic fields? 
 
Thermal emission from the neutron star itself is dominated by neutrinos because their interaction cross-section with
matter is weaker by $~10^{18}$ than that of high-frequency electromagnetic
radiation (the Thomson cross-section). While the photons are trapped
inside the neutron stars, the neutrinos have a much larger mean-free-path
and can diffuse out much faster. 
However, in a BNS coalescence  
the fraction of the heat carried out by photons may be much larger, because in this case the surface  
of the neutron stars rather than only the deep interior, as in supernova collapse, contribute. 
Thus we need to understand the ways of energy transport away from the neutron stars in different media. 
Moreover, the electromagnetic radiation might be dispersed in the interstellar medium, which may affect the observability of the 
memory effect. See \cite{bold1}. 

Unfortunately, the flux of neutrinos is too small to be detectable by
neutrino detectors from a source that resides outside our Milky-Way
galaxy. In a typical supernova, neutrinos carry as much as $10^{53} ergs$ 
away from the newly formed neutron star.  Even if tidal heating is weaker
for neutron star binaries, the neutrino losses may still dominate over
magnetic losses for moderate magnetic fields. However, 
magnetic losses might be more important for magnetars and BNS coalescence in low density medium. 

Observations during recent years have shown that BNS mergers also occur in media with lower electron density, for instance in elliptical galaxies or other galaxies that are gas poor. Moreover, according to recent observations and simulations some BNS mergers happen far away from the cores of galaxies due to the kick imparted to the two neutron stars by the supernovae that formed them. 
Compact binaries with kicks are investigated in \cite{bloom*2}, \cite{fryer}, \cite{belcz}. 
Another interesting point to mention here, which we will not pursue further, is that in certain binary black hole mergers the final object could be ejected from the host galaxy. See for instance \cite{loeb2} and references therein. 

Typical electron densities at the center of a giant elliptical galaxy are about $~ n = 0.1 $ per $cm^3$. This yields a plasma frequency of about $w_p = 20 rad/s \approx 3 kHz$ by (\ref{plasmafreq1}). Away from the center, the densities are much less, yielding plasma frequencies of significantly lower order. 
In intergalactic medium typical electron densities are $~ n = 10^{-6} / cm^3$. From (\ref{plasmafreq1}), we compute the corresponding  plasma frequency to be of the order $w_p = 60 rad/s \approx 10 Hz$. 
Summarizing, we find that electromagnetic waves can escape their origin as long as the surrounding medium has a plasma frequency which is much lower than magnitudes of $10 kHz$. 
Thus, in gas poor galaxies (such as elliptical galaxies) and far away from galactic centers, electromagnetic waves travel almost as in vacuum. 

These are therefore the places to look for BNS mergers producing an electromagnetic Christodoulou memory. As discussed in 
\cite{berger}, BNS coalescence is supposed to occur also in elliptical galaxies and far away from galactic cores, that is in media with low electron densities. 

In section \ref{bnsm} of this paper, we shall discuss these BNS mergers.

\section{Memory effect} \label{mem}

We cite the Bondi mass
loss formula obtained in \cite{zip2} for spacetimes solving EM equations: 
\begin{equation} \label{bondimassloss}
\frac{\partial }{\partial u}M\left( u\right) =\frac{1}{8\pi }%
\int_{S^{2}}\left( \left| \Xi \right| ^{2}+\frac{1}{2}\left| A_{F}\right|
^{2}\right) d\mu _{\overset{\circ }{\gamma }}.
\end{equation}
Compared to the formula obtained in \cite{sta} for spacetimes solving the EV
equations, we have an additional term, $|A_F|^2$, from the electromagnetic field.
See \cite{lpst1}. 

To study the effect of gravitational waves, 
we follow the method introduced in \cite{chrmemory}. 
The analysis is based on the asymptotic behavior of
the gravitational field obtained at null and spatial infinity. 
These rigorous
asymptotics allow to study the structure of spacetimes at null infinity. 
We foliate the spacetime by a 
time function $t$ and an optical function $u$, 
with corresponding lapse functions $\phi$ respectively $a$.  
Each level set of
$t$, $H_t$ is a maximal spacelike hypersurface, 
each level set of $u$, $C_u$, is an outgoing null hypersurface.  
Along the null hypersurface $C_u$, pick a suitable pair of normal
vectors. 
The flow along these vectorfields generates a
family of diffeomorphisms $\phi_u$ of $S^2$. 
Using $\phi_u$ we pull back tensor
fields. 
In this manner, we can study their limit at null infinity along
the null hypersurface $C_u$.  
Then take the limit as $u \to \pm \infty$, which allows 
to 
investigate the effect of gravitational waves. 
For a detailed explanation of the structure at null infinity, see \cite{chrmemory}. 
The following figure shows a schematic picture of an outgoing null hypersurface $C$, along which light travels, intersecting $H$ in $S$. 
The incoming null hypersurface is denoted by $\underline{C}$, the vectorfields $L$ and $\underline{L}$ defined below, lie in $C$ and $\underline{C}$ respectively. 

\begin{figure}[h]
\frame{\includegraphics[scale=0.6]{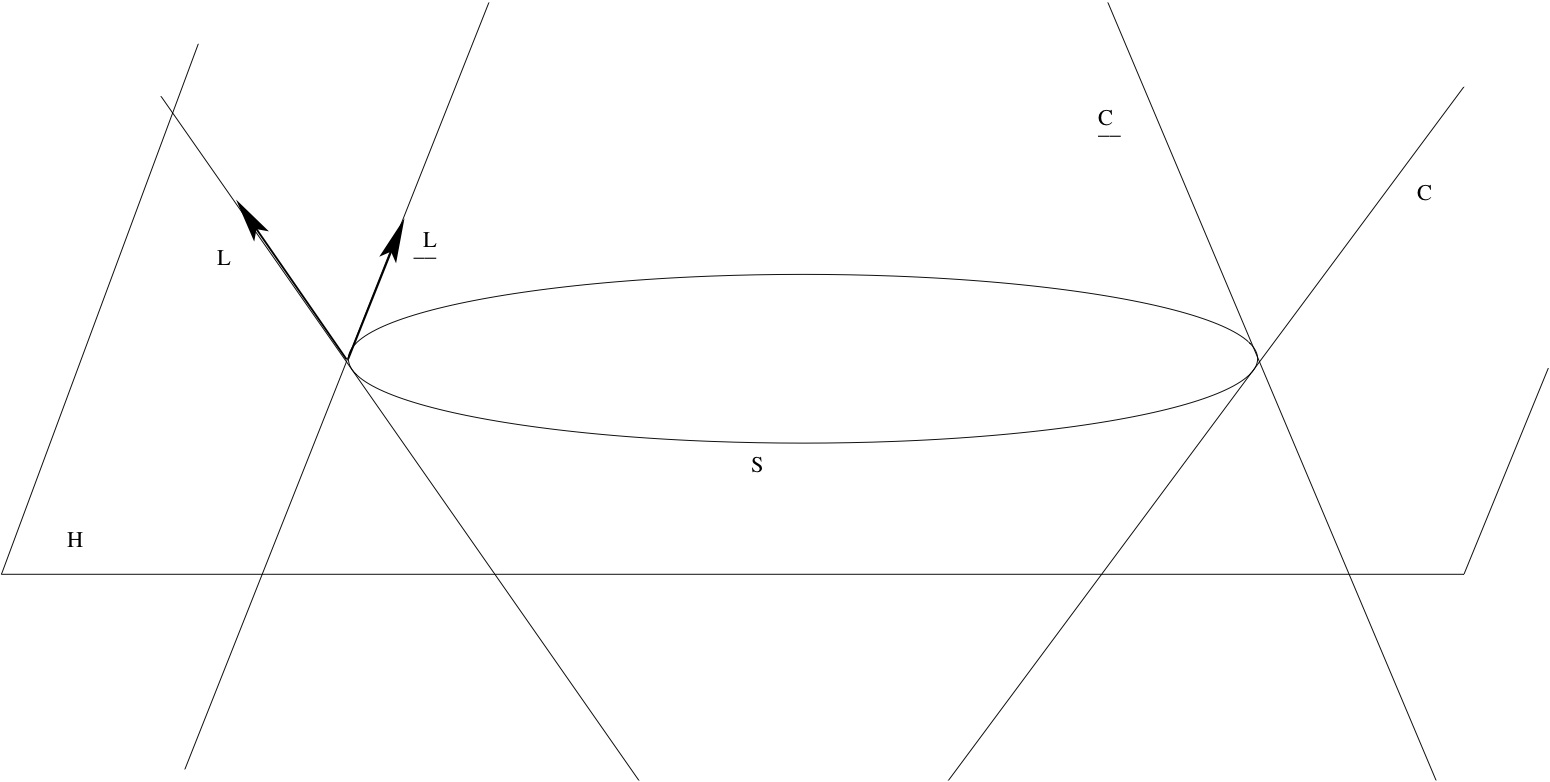}}
\caption{Schematic picture of an outgoing null hypersurface $C$, along which light travels, intersecting spacelike hypersurface $H$ in $S$. }
\end{figure}

The methods introduced in \cite{sta}, used in \cite{zip}, \cite{zip2} and
\cite{lydia1}, \cite{lydia2}, reveal the
structure of 
the null asymptotics of spacetimes. 
In these works, stability results were proven under smallness conditions on the initial data. 
However, the results about null infinity hold for large data.

In the Einstein-Maxwell equations the electromagnetic field is
represented by a 
skew-symmetric 2-tensor $F_{\mu\nu}$. The stress-energy tensor corresponding to
$F_{\mu\nu}$ is 
\[ T_{\mu \nu} = \frac{1}{4 \pi} \big(  F_{\mu}^{\ \rho} F_{\nu \rho} -
\frac{1}{4}g_{\mu \nu} F_{\rho \sigma} F^{\rho \sigma}  \big). \]
The Einstein-Maxwell equations read:
\begin{equation}
 \label{EM}
\begin{split}
\ R_{\mu \nu}  \ =& \ 8 \pi T_{\mu \nu}  \\
D^{\alpha} F_{\alpha \beta}  = & 0 \\
D^{\alpha} \ ^* F_{\alpha \beta} = & 0. 
\end{split}
\end{equation}
Indices run from $1$ to $4$. 
Let $S_{t,u}$ be the intersection of $H_t$ and $C_u$.
 Let $N$ be the spacelike unit normal vector of $S_{t,u}$ in $H_t$ and $T$ be the 
timelike unit normal vector of $H_t$ in the spacetime. Let $\{ e_a \}_{a=1,2}$  be
an orthonormal frame on $S_{t,u}$. 
Then 
$(T, N,e_2, e_1)$ is an orthogonal frame. 
This also gives us a pair of null normal vectors to $S_{t,u}$, namely $L=T+N$ and
$\underline{L}=T-N$. Together with $\{ e_a \}_{a=1,2}$, they 
form a null frame. 

We can decompose the Weyl curvature tensor and the electromagnetic field with
respect to either frame. The asymptotics of these
components are studied in 
\cite{zip} and \cite{zip2}. 
We will only list the components, which are used in this paper. 

Let $X, Y$ be arbitrary tangent vectors to $S$ at a point in $S$. 
Given the null frame $e_4 = L $, $e_3 = \underline{L}$ and  $\{ e_a \}_{a=1,2}$, let 
$\chi (X, Y) = g(\nabla_X L, Y)$ 
and $\underline{\chi} (X, Y) = g(\nabla_X \underline{L}, Y)$ be the second
fundamental forms with respect to $L$
and $\underline{L}$, respectively. Let $\widehat \chi$ and $ \widehat{\underline{\chi} }$ be their traceless parts.  
$\widehat \chi$ and $ \widehat{\underline{\chi} }$ are the shears. 
We also need the following null components of the Weyl curvature
\[ \underline \alpha_W(X,Y) = R(X, \underline L, Y , \underline L) \]
and the electromagnetic field 
\begin{equation}
\begin{array}{lll}
F_{A3}=\underline{\alpha }(F)_A  &  & F_{A4}=\alpha \left( F\right)_A
\\ 
F_{34}=2\rho \left( F\right) &  & F_{12}=\sigma \left( F\right) .
\end{array}
\label{null-electricfield}
\end{equation}
We have the following limits at null infinity
\[ \lim_{C_u , t \to \infty} r^2 \widehat \chi = \Sigma   \text{\, , \,}  \lim_{C_u
, t \to \infty} r \widehat{ \underline{\chi} }= 2 \Xi   \]  
\[ \lim_{C_u , t \to \infty} r \underline \alpha_W = A_W    \text{\, , \,}  \lim_{C_u , t \to \infty} r \underline{\alpha}_F = A_F.     \]
These are the main quantities in the description of gravitational radiation and memory.  
$\Sigma$ and $\Xi$ are the corresponding asymptotic shears, whereas $A_W$ and $A_F$ denote the asymptotic curvature and electromagnetic field components, respectively. 
As shown in \cite{chrmemory}, the permanent displacement of the test masses of a
laser interferometer gravitational-wave detector is 
governed by $\Sigma^+ - \Sigma^-$ where 
\begin{equation*} \lim_{u \to \pm \infty} \Sigma =  \Sigma ^{\pm}. \end{equation*}
In \cite{zip}, \cite{zip2} for the EM case were derived the following relations between the asymptotic shears $\Sigma$, $\Xi$ and the asymptotic curvature component $A_W$: 
\be \label{theXiSigmaA} 
\frac{\partial \Sigma}{\partial u} = - \Xi  \text{ \, and \,} \frac{\partial \Xi}{\partial u} = - \frac{1}{4} A_W. 
\ee
In \cite{lpst1}, we first write 
\be \label{Thm*FXiAF*1}
F (\cdot)  =   \int_{- \infty}^{\infty} 
\big( 
\mid \Xi (u, \cdot) \mid^2 + \frac{1}{2} \mid A_F (u, \cdot) \mid^2  
\big)
du 
\ee
where 
$\mathbf{\frac{F}{4 \pi}}$ is the total energy radiated to infinity in a given direction per unit solid angle. 
We then prove that 
in a spacetime solving the EM equations,
$\Sigma^+ - \Sigma^-$ is governed by the following relation. In particular, 
$\Sigma^+ - \Sigma^-$ is given by the following equation on $S^2$: 
\be \label{Thm*divSigma+-*2}
\stackrel{\circ}{\dlap} (\Sigma^+ - \Sigma^-) = \stackrel{\circ}{\nlap} \Phi 
\ee
where $\Phi$ is the solution with $\bar{\Phi} = 0$ on $S^2$ of the equation 
\[
\stackrel{\circ}{\slap} \Phi = F - \bar{F}  . 
\]
Comparing this with the EV case studied in \cite{sta} and used
in \cite{chrmemory}, 
where the corresponding formula was 
$F (\cdot)  =   \int_{- \infty}^{\infty} 
\mid \Xi (u, \cdot) \mid^2 du$, we find that new the electromagnetic part
$ \mid A_F (u, \cdot) \mid^2 /2$ appears in the integral. See \cite{lpst1}.  

\section{ Gravitational wave experiment} \label{gravitationalwaveexp}

How will our findings relate to experiment? We show how the 
electromagnetic field enters the gravitational wave experiments. In particular, we discuss the
instantaneous and the permanent displacement of test masses. 
For a detailed explanation of the experiment we refer to \cite{chrmemory} and for a
detailed derivation in the EM case we refer to \cite{lpst1}. 

Consider a laser interferometer gravitational-wave
detector with three test masses. 
We denote the reference mass by $m_0$, being also the location of the beam
splitter. 
The masses $m_1$ and $m_2$ are initially at right angles and equal distances $d_0$ from $m_0$. 
The masses are free if the observation is performed in space 
as in LISA, or are suspended by pendulums if the experiment is performed on the Earth as in LIGO. 
In the second case, for time scales much shorter than the period of the pendulums the motion 
of the masses in the horizontal plane can be considered free. 
One measures the distance of  $m_1$ and $m_2$ from $m_0$ by laser interferometry. 
We observe a difference of phase of the laser light at $m_0$ whenever the light
travel times between $m_0$ and $m_1$, $m_2$, respectively, differ. 

The motion of the masses $m_0$, $m_1$, $m_2$ is described by geodesics $\gamma_0$,
$\gamma_1$, $\gamma_2$ in spacetime. 
Denote by $T$ the unit futuredirected tangent vectorfield of $\gamma_0$ and by $t$
the arch length along $\gamma_0$. 
Let $H_t$ be for each $t$ the spacelike, geodesic hyperplane through $\gamma_0
(t)$ orthogonal to $T$. 
At $\gamma_0 (0)$ pick an orthonormal frame $(E_1, E_2, E_3)$ for $H_0$. By
parallelly propagating it along $\gamma_0$, 
we obtain the orthonormal frame field 
$(T, E_1, E_2, E_3)$ along $\gamma_0$, where at each $t$ the $(E_1, E_2, E_3)$ is an
orthonormal frame for $H_t$ at $\gamma_0 (t)$. 
Then we assign to a point $p$ in spacetime close to $\gamma_0$ and lying in
$H_t$ the cylindrical normal coordinates $(t, x^1, x^2, x^3)$.  
We shall assume that the source of the waves 
is in the direction $E_3$. 

Suppose that the light travel time corresponding to the distance $d_0$ is much shorter than the time scale 
in which the spacetime curvature varies significantly. 
Then 
the geodesic deviation from $\gamma_0$, namely the Jacobi equation (\ref{jacobi*1}),
replaces the geodesic equation for $\gamma_1$ and $\gamma_2$. 
Let $R_{k0l0} = R(E_k, T, E_l, T)$, then 
\be \label{jacobi*1}
 \frac{d^2 x^k}{dt^2}=-R_{k0l0} x^l .
 \ee
 On the left hand side of (\ref{jacobi*1}) we find an acceleration term, controlled by the curvature $R_{k0l0}$ on the right hand side. 
 Now, we have to lay open the structure of this curvature term to derive the contributions from gravitational and electromagnetic parts. 
Decompose $R_{k0l0}$ into Weyl curvature and Ricci curvature
\[ R_{k0l0}=W_{k0l0} +
\frac{1}{2}(g_{kl}R_{00}+g_{00}R_{kl}-g_{0l}R_{k0}-g_{0k}R_{l0}). \]
From the EM equations (\ref{EM}) we find 
\be \label{R00}
R_{00} =  \frac{1}{2} ( \mid \underline{\alpha} (F) \mid^2 + \mid \alpha (F) \mid^2 
)  + \rho (F)^2 + \sigma (F)^2 .
\ee 
The component $R_{00}$ observes the term $\mid \underline{\alpha} (F) \mid^2$, where
$\underline{\alpha} (F)$ is the electromagnetic field component with worst decay
behavior, but entering $R_{00}$ as a quadratic. 
Hence, $R_{00}$ is of order $O(r^{-2})$. Whereas the leading order component of
the Weyl curvature is of order $O(r^{-1})$. 
(See \cite{lpst1})  
Thus, the electromagnetic field does not contribute at highest order  \footnote{Here and in what follows, the word `order' refers to decay behavior of the exact solution, not to any approximations. That is, `higher order' means `less decay'. For details, see \cite{lpst1}. \cite{lydia2}, \cite{zip2}.} to the
deviation measured by the Jacobi equation. 
As a consequence, it does only change at lower order the instantaneous displacement
of test masses. 
However, it changes the nonlinear memory effect: 
Using (\ref{theXiSigmaA}) and (\ref{Thm*FXiAF*1}) as
well as the fact that $\Xi \to 0$ for $u \to \infty$ and taking the limit 
$t \to \infty$, we conclude that the test masses experience permanent displacements
after the passage of a wave train. In particular, 
this overall displacement of test masses is described by $\Sigma^+ - \Sigma^-$ 
\be \label{overall} 
\Delta x^A_{(B)} = - \frac{d_0}{r}(\Sigma^+_{AB}-\Sigma^-_{AB}) 
\ee
where from ((\ref{Thm*FXiAF*1}), (\ref{Thm*divSigma+-*2})) one sees that the right hand side of
(\ref{overall}) includes the electromagnetic field terms at highest order. 

To derive (\ref{overall}), we use $L = T - E_3$ and
$\underline{L} = T + E_3$. 
We have the leading components of the curvature $\underline{\alpha}_{AB} (W)$
and of the 
electromagnetic field $\underline{\alpha}_A (F)$: 
\beas
\underline{\alpha}_{AB} (W) & = &  R  (E_A, \ \underline{L},  E_B, 
\underline{L})  =  \frac{A_{AB}(W)}{r}  +  o  (r^{-2})  \\ 
\underline{\alpha}_{A} (F) & = &  F (E_A, \underline{L})   = 
\frac{A_{A}(F)}{r}  +  o  (r^{-2}) .
\eeas

Let $x^k_{(A)}$ with $A = 1,2$, denote the $k^{th}$ Cartesian coordinate of mass
$m_A$. 
From \cite{chrmemory} and \cite{lpst1} one sees that there is no acceleration to
leading order in the vertical direction. 
Start with $m_1$, $m_2$ being at rest at equal distance $d_0$ from $m_0$ at
right angles from $m_0$. 
Thus to leading order it is 
\be 
\ddot{x}^A_{(B)}  =  - \frac{1}{4} r^{-1} d_0 A_{AB} .
\ee
In particular, the initial conditions are as $t \to - \infty$: \\ 
$x^B_{\ (A)} =  d_0  \delta^B_A  ,  
\dot{x}^B_{\ (A)}  =  0 , 
x^3_{\ (A)}  =  0  ,  \dot{x}^3_{\ (A)}  = 0$. 

Integrating gives 
\be 
\dot{x}^A_{\ (B)}  (t)  =  -  \frac{1}{4} \  d_0   r^{-1}  
\int_{- \infty}^t \ A_{AB} (u)  d u .
\ee
From (\ref{theXiSigmaA}) equation $ \partial \Xi/\partial u = -
\frac{1}{4} A_W $ and 
$lim_{\mid u \mid \to \infty} \Xi = 0$, one substitutes and concludes 
\be
\dot{x}^A_{\ (B)} \ (t)  =  \frac{d_0}{r}   \Xi_{AB} \ (t)  .
\ee
As $\Xi \to 0$ for $u \to \infty$, the test masses return to rest after 
the passage of the gravitational waves. 
Use (\ref{theXiSigmaA}) equation $\partial \Sigma/\partial u =
- \Xi$ and 
integrate again to obtain 
\be
x^A_{\ (B)}  (t)  =  -  (\frac{d_0}{r})  (\Sigma_{AB}  (t)  -  \Sigma^-) .
\ee
Finally, by taking the limit $t \to \infty$ one derives that the test masses obey
permanent displacements. 
This means that 
$\Sigma^+ - \Sigma^-$ is equivalent to an overall displacement of the test masses
given by (\ref{overall}):
\be \label{permdispl*}
\triangle  x^A_{\ (B)}  =  -  (\frac{d_0}{r})  (\Sigma^+_{AB}  - 
\Sigma^-_{AB}) .
\ee
The right hand side of (\ref{overall}) includes terms from the
electromagnetic field at highest order as given in our relations (\ref{Thm*FXiAF*1}), (\ref{Thm*divSigma+-*2}). 

\begin{figure}[h]
\frame{\includegraphics[scale=0.5]{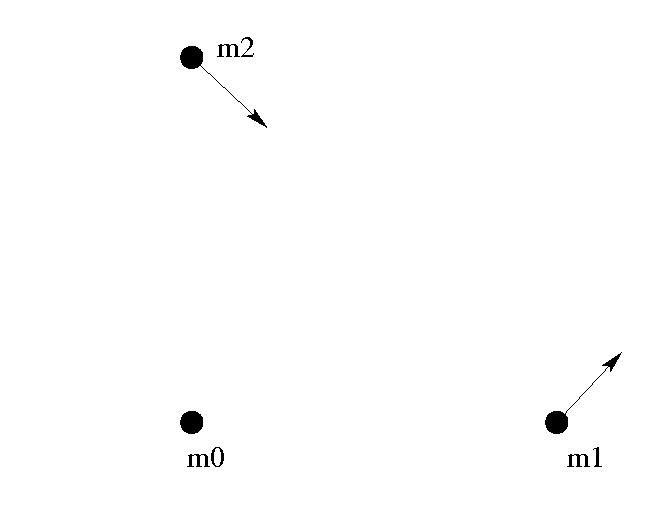}} 
\caption{Permanent displacement of test masses caused by Christodoulou memory effect. Test masses $m1$ and $m2$ are displaced permanently after the passage of a gravitational wave train.}
\end{figure}

The magnitude of the term $\mid \Sigma^+_{AB}  - \Sigma^-_{AB} \mid$ on the right hand side of (\ref{permdispl*}) 
is of the order of the total energy radiated during the merger of the source objects. 
The nonlinear effect builds up over the time scale when the energy is radiated. 
An experiment like LIGO for instance would measure differences in distances $\triangle x$ of test masses, 
whereas the radio telescopes would measure changes in the frequency of pulsars' pulses. 
The reader may want to compare our formulas (\ref{permdispl*}) and (\ref{Thm*FXiAF*1}) with the corresponding result in 
\cite{chrmemory} as well as with computations using Post-Newtonian (PN) approximation for instance in \cite{favata9} or recent results in \cite{favata99} and references therein. 

In particular, how does our result compare to the pioneering paper by \cite{chrmemory}? In order to answer this question in detail, let us cite 
from  \cite{chrmemory} p. 1488: ``When matter (i.e., electromagnetic or neutrino) radiation is present then if $T$ is the energy tensor of matter, 
$\phi^*_u (r^2 \frac{1}{4} T(\underline{l}, \underline{l}))$ tends to a limit $E$ as $r_0^* \to \infty$ and in (7)-(9) $\mid \Xi \mid^2$ is replaced by 
$\mid \Xi \mid^2 + 32 \pi E$." This is a suggestion, in which direction one would have to search to find other contributions to the nonlinear memory effect. It was not known, what the limit $E$ would be. This limit $E$ depending on $u$ could behave in such a way that 
there were no additive contribution from $E$ to the memory, or that it was negligible. 
Studying the adapted formulas (7)-(9) in \cite{chrmemory}, one has to keep in mind that formula (9) governs the nonlinear memory effect. It is an additive effect. 
How do we know that $E$ is in fact contributing? What is the structure of this limit? We give the answer in our formulas (\ref{permdispl*}) and (\ref{Thm*FXiAF*1}) based on \cite{lpst1} and on (\ref{bondimassloss}) from \cite{zip2}. 
Our formula (\ref{Thm*FXiAF*1}) corresponds to Christodoulou's formula (9). 
We find that the limit $A_F$ has the same decay behavior in $u$ as the limit $\Xi$. Namely they satisfy 
\beas
\mid A_F (u, \cdot) \mid & \leq & C_1 (1 + \mid u \mid)^{- \frac{3}{2}} \\ 
\mid \Xi (u, \cdot) \mid & \leq & C_2 (1 + \mid u \mid)^{- \frac{3}{2}} 
\eeas

Knowing these structure, we investigate our formula (\ref{Thm*FXiAF*1}) more closely. 
Integrating with respect to $u$ from $- \infty$ to $+ \infty$ yields a positive constant for $F$. This value contains the corresponding positive 
constants coming from the electromagnetic field term $A_F$ and from the purely gravitational term $\Xi$. 
This proves that the contribution from the electromagnetic field is of the same order  \footnote{Here, the word `order' refers to decay behavior of the exact solution, not to any approximations. That is, `higher order' means `less decay'. For details, see \cite{lpst1}. \cite{lydia2}, \cite{zip2}.} as the purely geometric part. 
Our result being exact, it holds for all corresponding physical situations. 
The constants $C_1$ and $C_2$  have to be determined or estimated from astrophysical data of the many scenarios. This will be the purpose of the following section, where we give rough estimates. It will be a challenge for the future to work on the many details. 

Summarizing, we have in (\ref{Thm*FXiAF*1}) a general formula that always holds. Thus we can apply it to all situations. From astrophysical data we can now determine the corresponding contributions in every scenario. 

Again one could ask, what if one linearized the equations and just looked at the linear part in the approximation. We have explained in the introduction, why that produces a physical error. However, for the purpose of discussion, let us consider it for a moment. 
The limit $A_F$ in our fomulas (\ref{permdispl*}) and (\ref{Thm*FXiAF*1}) or in (\ref{bondimassloss}) is exact. Solving the full nonlinear problem yields our exact formulas revealing the precise physics. However, in a linearized version, one would only obtain a formula that `looks like' ours, but does not capture the physics. In general, for any energy momentum tensor $T$ mentioned previously this linear part is expected to be different from the exact solution, and therefore the true physical picture gets lost when linearizing.

\section{ Binary neutron star mergers} \label{bnsm}

We compute the electromagnetic Christodoulou memory effect for typical sources, that
is for different constellations of BNS mergers as stipulated in section \ref{ns}. 
For an overview on compact neutron star binaries, see for instance \cite{duez}. 

In this section, the data is taken from different papers we are citing and which suggest different values for 
the same quantities. Combining these findings, it gives certain ranges of magnitudes for the corresponding data. 

In a binary neutron star or binary black hole system, the two objects are orbiting
each other. In Newtonian physics, they would stay like that forever. However,
according to the theory of general relativity such a system must radiate away
energy. Therefore, the radius of the orbits must shrink and finally the objects will
merge. 
The magnetic fields generated and radiated during the merger of two neutron
stars are among the largest magnetic fields 
known in astrophysics. In fact, in the electromagnetic Christodoulou memory effect
that we derived, the magnetic field enlarges the 
nonlinear displacement of (non-charged) test masses significantly. 

Observations and numerical simulations suggest that magnetic fields of neutron stars may take values between 
$10^{12} G$ and $10^{16} G$, the latter being the fields of magnetars. See \cite{duncanandthompson}. 
To start with, one can consider the merger of two magnetars, even though this is expected to be a very rare (or impossible) event. Calculations with large magnetic field strengths such as in magnetars have been carried out in the GR regime in 
\cite{anderson1}. 
On the other hand, in \cite{pricerosswog} in a Newtonian regime, it has been shown that moderate magnetic fields for ordinary neutron stars, thus of the order $B \approx 10^{12} - 10^{13} G$ rapidly grow during merger. Very strong magnetic fields could occur in the early shear phase and afterwards due to dynamo type effects. The magnetic field can reach up to $ 10^{16} G$. 
It had also been observed that magnetic fields delay the merger time. 
See also \cite{lset} and \cite{grb1}. 
In \cite{grb1} the values for the magnetic fields are found to be smaller. The paper \cite{pricerosswog} also mentions magnetized blasts as well as processes in the accretion torus with buoyant magnetic fields at about $10^{14} G - 10^{16} G$. 

We conclude that the electromagnetic Christodulou memory should be very important 
in the coalescence of BNS (also with moderate initial magnetic fields) in gas poor galaxies and far away from galactic cores. In these processes, the magnetic field will grow to values of $B \approx 10^{16} G$. Moreover, the energy radiated by electromagnetic waves may dominate the radiation by neutrinos in these situations. 
We will analyze these points in the remainder of this article. 

Astrophysical data give for typical neutron star binaries a range of possible
constellations which allow the mass and the 
magnetic field to vary within given boundaries \cite{st}, \cite{duez} and references therein. Typically, the mass of a neutron
star is around slightly more than $1 M_{\sun}$ and the radius 
of a neutron star is $5$ - $ 16$ km. Thus, the typical mass for a BNS system ranges
between $2.6$ and $2.8 M_{\sun}$ \footnote[1]{$ M_{\sun} = 1$ solar mass $ \approx
1.9891 \cdot 10^{33}$ g}. 
In such a system, the neutron stars are spiraling 
around each other, radiating neutrinos, gravitational and magnetic energy. 
The inspiral gains speed and the BNS system emits an increasing
amount of electromagnetic and gravitational energy, which becomes extremely large
when the orbit radius is about $10$ - $100 $km.  
For the detection of the electromagnetic Christodoulou effect, the largest
contribution will come from the last phase of the inspiral, starting when the orbit
radius is about 10 times the neutron star radius. 
In the literature, we find that the merger and following collapse times range from a few milli-seconds up
to $1000$ ms. We compare the amount of 
gravitational energy radiated during the merger to the amount of magnetic
energy radiated. On the one hand, the amount 
of gravitational energy is well-known. In general, about $1 \%$ of the
initial mass is radiated during a merger. This is about 
$10^{52}$ erg.\footnote{$1 \  {\rm erg} = 1 {\rm g \cdot cm^2s^{-2}}$ and $1 M_{\sun}
\approx 1.78 \cdot 10^{54}$ erg} On the other hand, the amount of magnetic energy radiated 
could vary 
drastically depending on different constellations. Typically, the rate of change for
the magnetic field is $dB/dt \approx 10^9 - 10^{17} \,{\rm G(ms)^{-1}}$ and the
magnetic field produced in 
the merger is about $10^9 - 10^{17} \,{\rm G}. \footnote{$1$ Gauss: $1 G = 10^{-4}\, {\rm kg \cdot
C^{-1} s^{-1}} = 10^{-1}\,{\rm g \cdot C^{-1}  s^{-1}}$  }$ 
(For these ranges, see references given in this section. The values in the different papers vary.)

Assume: Total mass of BNS is initially $ 2 M_{\sun}$, $1 \%$ of the total mass will
be radiated away during the (whole) merger, the radius of each neutron star is $10$ km.
Under these assumptions, the gravitational energy radiated away is about $3.56 \times 10^{52} $\,erg. 
In the first example to follow, we will assume that the full initial magnetic field will be radiated, whereas in the second example the magnetic field will increase during merger and be radiated, yielding a ratio of $B_{radiated} / B_{initial} > 1$. 

In the physics literature, one finds many linearized models. However, since the Einstein
equations are nonlinear, the main information usually gets lost in linearized
models. As we do investigate the nonlinear problem here, and as 
the results of \cite{chrmemory} and \cite{lpst1} show the Christodoulou memory
effect of gravitational waves to be a nonlinear phenomenon, we consider a
corresponding nonlinear model for the neutron star binary coalescence. 
Thus, we use the results of Zipser's global stability work \cite{zip} and
\cite{zip2} for the 
initial value problem in 
spacetimes satisfying the Einstein-Maxwell equations. 
We assume that outside the neutron star, the magnetic field decays like $r^{-5/2}$.
Such decay at spatial infinity is suggested by the decay obtained in \cite{zip},
\cite{zip2}. 
One might want to consider situations with a slightly different decay of the
magnetic field. This would not affect the main picture, as one finds during the
computations that the decay of the magnetic field does not play a role here.  \footnote{This is not to be confused with the findings in the previous sections. Whereas the structure of the magnetic fields, in particular the decay behavior with respect to $r$ and $u$, is crucial to determine if a nonlinear memory exists, in the present situation, we want to estimate the size of the corresponding energies, that is constants in the formulas.} 
In the following paragraphs, we are going to present two scenarios of BNS coalescence. 

Consider such a BNS system with 
the magnetic field $B$ 
initially being $B = 10^{13}$ G and $dB/dt = 10^{13}\, {\rm G(ms)^{-1}}$ on the
surface of the neutron star. 
Assume that the merger time is $50$ ms. We estimate the total magnetic energy
radiated using the following model. We assume that throughout the merger, the
matter of the neutron star
stays in a ball of radius $10$ km. We compute the contribution from the 
magnetic field outside 
the support of the matter of the neutron star. As a result, we simply use the vacuum
magnetic constant when computing
the magnetic energy density. 
Moreover, we assume that outside the neutron star, the magnetic field decays at the rate of 
$r^{-5/2}$.
Using this model, the energy radiated from the 
magnetic field is about $1.19 \cdot 10^{47}$ erg. In this case, the addition
of a magnetic field has a small contribution to the memory effect. 
This is similar to the 
numeric simulation in  \cite{grb1}. Their result suggests that for an initial magnetic field smaller than  $10^{14}$ G, 
the gravitational wavefront does not change much.

Next, consider a BNS system with the above data, but where the magnetic field $B$ is initially $B = 10^{15}$\,G and $dB/dt = 10^{16}\, {\rm G(ms)^{-1}}$ on the surface of the neutron star. Assume that the merger time is $50$ ms. We compute that the total magnetic energy radiated is about $1.19 \cdot 10^{53}$ erg. This is higher than the
gravitational energy radiated. This situation is consistent with astrophysical data.  
This example here is similar to the numeric simulation in \cite{anderson1}, where an initial $B = 9 \times 10^{15}$ is used.
In \cite{anderson1}, the merger time is of the order of 10 ms and it is noticed that 
the gravitational-wave signal is considerably stronger and the merger could be detected by LIGO at twice as far away with the addition of a magnetic field. While the largest magnetic field during merger is 
not explicitly given there, from the numeric simulations in \cite{grb1}, \cite{grb2} and \cite{grb3}, it is observed that the magnetic field could increase by two orders of magnitude
during merger when one starts with magnetic fields around $10^9$ to $10^{12}$ G. When we start with a stronger magnetic field, the merger
would take longer and allow more time for the magnetic field to build up. Also magnetic fields of similar initial strength ( $B = 10^{15}$\,G) are used in the simulation of \cite{lset}. 
Their simulations suggest that the addition of the magnetic field causes observable differences in the dynamics and gravitational waveforms and that the most important role of magnetic fields are on the long term evolution. This is similar to our conclusion from (\ref{theXiSigmaA}) and (\ref{Thm*FXiAF*1}), (\ref{Thm*divSigma+-*2}). Namely, the addition of a magnetic field does not change the system instantaneously but it does contribute to the nonlinear long-term permanent change. While our initial magnetic field is rather large, it could still be realized by sources such as magnetars. Also, as noted in \cite{pricerosswog}, the outflow of energy driven by magnetic mechanisms could easily reach $10^{51}$ ergs or more. 
Moreover, according to \cite{anderson1} and \cite{pricerosswog}, BNS mergers with initial magnetic fields of ordinary neutron stars, namely of the order $B \approx 10^{13} G$, produce very large magnetic fields. This yields magnitudes for $dB/dt$ of the order 
$dB/dt \approx 10^{16}\, {\rm G(ms)^{-1}}$. 
These huge magnetic fields are expected to occur at the shear layer during merger and during the subsequent collapse of the merger product into a black hole by dynamo effects. 

To compare, note that the amount of energy emitted in a binary black hole merger 
is expected to be as follows: a binary black hole system with equal mass and no spin
would lose about $4\%$ of the mass during merger. 
However, according to the present knowledge in astrophysics, BNS mergers occur more often than black hole mergers. 

{\bf Conclusions:} We find that the best conditions for magnetic fields to increase the Christodoulou memory 
occur in BNS mergers in medium with a plasma frequency of less than $30 kHz$, that is 
in gas poor regions (for instance elliptical galaxies) and far away from galactic centers. 
Since in these regions, the energy carried away by neutrinos is not expected to dominate the electromagnetic radiation, 
the latter will play an important role. In fact, the magnetic fields of typical sources contribute to the Christodoulou memory at the same highest order  as the purely gravitational term. 
The merging objects can be ordinary neutron stars with initial magnetic fields of the order $B \approx 10^{13} G$. These fields will be enlarged during the merger process and subsequent collapse to a black hole, yielding $dB/dt \approx 10^{16} G$. Thus, magnetic fields of these orders radiated through a very short time will enlarge the Christodoulou memory considerably. 
We have also found that the amounts of energy radiated in BNS mergers of this type are expected to be $10^{52} ergs$ in form of gravitational waves, about $10^{47} ergs - 10^{53} ergs$ in electromagnetic waves and about $10^{53} ergs$ in form of neutrinos. We stress the fact that these are `only' energy considerations. It has not yet been shown that the neutrinos would produce a memory effect as well. Even though this might look plausible, it still has to be investigated. So far, two of the three main ways of energy transport in a BNS merger, namely gravitational and electromagnetic, have been shown to yield a Christodoulou memory effect. 

{\bf Acknowledgment:} We are very grateful to Demetrios Christodoulou, to Avi Loeb, to Bence Kocsis and to Chris Miller.  
We thank Demetrios Christodoulou for sharing his valuable insights with us,  
for fruitful discussions and
his interest in this work. We thank Avi Loeb for his valuable comments and for pointing out to us many of the 
astrophysical backgrounds and references. 
We thank Bence Kocsis for his valuable feedback and comments. We thank Chris Miller for his valuable feedback and comments. 
L. Bieri is supported by NSF grant DMS-0904583 
and S.-T. Yau is supported by NSF grant PHY-0937443 and DMS-0904583. \\ \\ \\

\end{document}